\documentclass[aps,prl,reprint,floatfix,twocolumn]{revtex4-2}

\usepackage[utf8]{inputenc}
\usepackage[english]{babel}          
\usepackage{amsmath,amsthm,amsfonts,amssymb,amscd}
\usepackage{physics}
\usepackage{graphics,graphicx,latexsym}
\usepackage{enumerate}
\usepackage[caption=false]{subfig}   
\usepackage[colorinlistoftodos,portuguese,textwidth=140pt]{todonotes}
\usepackage{hyperref}
\usepackage{mathtools}
\usepackage{booktabs}
\usepackage{caption}                 

\begin{document}

\title{Optimal beam displacement measurements using high-order structured light modes}

\author{A. L. S. Santos Junior$^{1}$, J. C. de Carvalho Junior$^{1}$, M. Gil de Oliveira$^{1}$, \\
E. V. S. Cubas$^{1}$, R. F. Barros$^{2}$, A. Z. Khoury$^{1}$, G. B. Alves$^{1}$}

\affiliation{$^{1}$Instituto de Física, Universidade Federal Fluminense, CEP 24210-346, Niterói-RJ, Brazil\\
             $^{2}$Instituto de Física, Universidade de São Paulo, CEP 05508-090, São Paulo-SP, Brazil}

\date{\today}

\begin{abstract}
We develop a novel technique to measure small angular and lateral displacements of
structured light beams. The technique relies on using high-order Hermite-Gaussian (HG) and Laguerre-Gaussian (LG) modes, which have well-defined symmetry under inversion. We show that the displacements of such fields lead to a crosstalk with modes with opposite parity under inversion, which we measure optimally with an interferometric parity sorter. Using this technique, we achieve improvement factors of up 41 in the signal-to-noise ratio using Hermite-Gaussian
modes and 21 using Laguerre-Gaussian modes with order up to 20, as compared to the fundamental Gaussian mode. Our results present a viable way of using structured light for metrology that
does not demand quantum light or homodyne detection.
\end{abstract}

\maketitle

\textit{Introduction.}---The standard way to measure a small optical angular displacement is using a lens followed by a split detector positioned in the focal plane of the lens \cite{putman}. This simple arrangement is capable of reaching around 64\% of the maximum theoretical precision \cite{knee_amplification}, however it is only applicable to a beam in the fundamental Gaussian mode. One way to overcome this limitation is by using spatially multimode light in which one of the modes is in a squeezed detection mode \cite{PhysRevLett.88.203601,quantum_laser_pointer,treps_nano-displacement_2004}. A more efficient way to measure small displacements in a Gaussian mode of a laser beam is to measure the first order Hermite-Gaussian mode ($HG_{1,0}$) of the displaced beam, which can be achieved through homodyne detection in that particular mode \cite{hsu_optimal_2004,Delaubert:06,PhysRevA.74.053823}. In order to beat the standard quantum noise limit (SQL), it is necessary to prepare the probe beam with a $HG_{1,0}$ squeezed mode, where improvements of 2 dB were obtained for displacements measurements \cite{PhysRevA.74.053823,Delaubert:06}. Other techniques involving weak-value amplification were also developed \cite{spin_hall_wv,ultrasensitive_beam_deflection,optimizing_signal_to_noise}, where the amplification effect comes at a cost of reduced statistics of the post-selected events, which may limit the overall precision \cite{PhysRevA.89.052117,PhysRevLett.112.040406}, although it can asymptotically reach the ultimate precision limit with a judicious post-selection \cite{PhysRevA.85.060102,PhysRevA.91.062107,PhysRevA.95.012104}. However, all these schemes consider the probe state in a Gaussian mode.

A more general situation can be considered when the probe laser beam is prepared in a high-order Hermite-Gaussian ($HG_{m,0}$) mode. In this case, the minimum measurable displacement in the horizontal direction is reduced by a factor of $\sqrt{2m+1}$ \cite{Gao-APL}. For that, a homodyne measurement is required in which the local oscillator is in a superposition of $HG_{m-1,0}$ and $HG_{m+1,0}$. An improvement factor of $\sim 3$ dB was obtained in the signal-to-noise (SNR) when the probe was in the $HG_{1,0}$ mode and the local oscillator in a suboptimal mode $HG_{2,0}$ \cite{Gao-APL}. Recently, it was possible to prepare the optimal local oscillator for the probe modes up to $HG_{4,0}$, yielding an enhancement by $9.2$ dB in the SNR \cite{photonics10050584}. Moreover, a weak-value-based measurement can also be conceived which provides the same $\sqrt{2m+1}$ improvement factor \cite{PhysRevApplied.13.034023}. It is worth noting that quantum states of light bring metrological advantages to the measurement, such as the squeezed modes $HG_{m-1,0}$ / $HG_{m+1,0}$, however it is a challenging task to produce such quantum states, which are still very fragile to losses. Nonetheless, an enhancement of 10 dB and 8.6 dB were obtained in the SNR for tilt and displacement measurements, respectively, with a squeezed fourth-order Hermite-Gaussian beam \cite{Gao-AdvPhotonics2022}, yielding an improvement factor of 3.2 and 2.7, respectively, in the measurement.

The measurement of an optical beam displacement is essential for many technological applications, such as atomic force microscopy \cite{putman,santhanakrishnan_non-contact_1995}, optical tweezers \cite{SIMMONS19961813}, satellite alignment \cite{Arnon:98}, and even for biological purposes as for single molecule tracking \cite{Denk:90,kojima_mechanics_1997,taylor_biological_2013}. Moreover, the use of structured modes has significant applications in various research areas, including those related to metrology. For instance, the use of optical tweezers with Laguerre-Gaussian (LG) modes or superpositions of modes has proven advantageous in the manipulation of trapped particles \cite{yang2021optical, otte2020optical, melo2020optical}. Additionally, the enhanced capacity for information encoding in structured modes is particularly appealing for free-space communications \cite{krenn2014communication, du2015high}. Similarly, the employment of higher-order Hermite-Gaussian (HG) modes has demonstrated great potential in reducing thermal noise in gravitational wave detectors \cite{ast2021higher}. All these applications demand the sensitive measurement of the optical beam displacement.

In this letter, we present a novel method to measure displacements of Hermite-Gaussian and Laguerre-Gaussian modes with optimal sensitivity. Our method relies on an interferometer design capable of sorting light fields by their parity under inversion symmetry. We show that for HG and LG modes, small displacements 
translate into a crosstalk with modes of opposite parity, which can be measured optimally in our experiment. Using this technique, we achieve optimal metrological performance with a single structured light beam, without the need for a local oscillator and homodyne detection.

\textit{Theory.}---A lateral displacement $d$ in the $\hat{x}$ direction in a laser beam can be seen as an unitary operator given by $\exp(-id\hat{p}_x)$, where
\begin{equation}
  \hat{p}_x=-i\frac{\partial}{\partial x}\,.
\end{equation}
Here, we consider the case of a small displacement with respect to the width of the mode, so that we can restrict to the first-order approximation:
\begin{equation}\label{eq-approx-lateral}
  e^{-id\hat{p}_x}|\Psi\rangle\approx(1-id\hat{p}_x)|\Psi\rangle\,.
\end{equation}
If the mode $|\Psi\rangle$ has a well-defined parity under the transformation $\Vec{r}\to-\Vec{r}$, we can see that the first and second term in the expansion \eqref{eq-approx-lateral} are orthogonal. Therefore, we see that the full information about the displacement ($d$) is obtained by measuring the component orthogonal to the original state $|\Psi\rangle$.

Two notable examples of modes with well-defined parity are the Hermite and Laguerre-Gaussian modes. Consider a coherent state of a laser beam in a collimated Hermite-Gaussian ($HG_{m,0}$) or in a Laguerre-Gaussian ($LG_{m,n}$) spatial mode propagating in the $z$ direction:
\begin{align}
    &HG_{m,0}(x,y,z)=C_m\,H_m\left(\frac{\sqrt{2}x}{w}\right)e^{-\frac{(x^2+y^2)}{w^2}}e^{ikz}\,, \label{Hermite-mode}\\
    &LG_{m,n}(r,\phi,z)=C_{m,n}\,\left(\frac{\sqrt{2}r}{w}\right)^{|m-n|}(-1)^{\min(m,n)} \nonumber\\
    &\times  L^{|m-n|}_{\min(m,n)}\left(\frac{2r^2}{w^2}\right)e^{-\frac{r^2}{w^2}}e^{-i(n-m)\phi}e^{ikz}\,,
\end{align}
where $H_m(x)$ is the Hermite polynomial of order $m$, $L^l_p(x)$ is the generalized Laguerre polynomial, $w$ is the beam width ($1/e^2$ width for the fundamental Gaussian mode), $k$ is the wave number and $C_m=\sqrt{\tfrac{2\times 2^{-m}}{\pi w^2 m!}}$, $C_{m,n}=\sqrt{\tfrac{2}{\pi w^2 m!n!}}\min(m,n)!$ are normalization constants. The mode order is given by the sum of the two label indexes. This indexes $m$ and $n$ for the Laguerre-Gaussian modes can be converted onto the usual orbital angular momentum ($l$) and radial index ($p$) with the following relations: $l=n-m$ and $p=\min(n,m)$ \cite{BEIJERSBERGEN1993123}.

Let us first consider the case of a displaced Hermite-Gauss beam. Under a small tilt ($d/w\ll 1$), the beam can be expanded in the original $HG$ basis as
\begin{align}\label{mode-expansion-lateral}
    &e^{-id\hat{p}_x}\ket{HG_{m,0}}\approx (1-id\hat{p}_x)\ket{HG_{m,0}}= \ket{HG_{m,0}} \nonumber\\
    &+\frac{d}{w}\left(\sqrt{m+1}\,\ket{HG_{m+1,0}}-\sqrt{m}\,\ket{HG_{m-1,0}}\right)\,,
\end{align}
for $m\geq 1$ and neglecting second-order terms in the series expansion \cite{supplemental-material}.

The lower bound for the uncertainty on the estimation of the parameter $d$, $\delta d$, is provided by the Cramér-Rao bound which establishes that $\delta d\geq 1/\sqrt{F(g)}$, where $F(g)$ is the Fisher information associated to the set of probabilities $P_i(d)$ of a given measurement, such that $F(d)=\sum_i 1/P_i(d)\left[\partial P_i(d)/\partial d\right]^2$. The maximization of the Fisher information over all possible quantum measurements leads to the \textit{quantum Fisher information}, and for a unitary operation, its value is given by $\mathcal{F}=4(\langle\hat{H}^2\rangle-\langle\hat{H}\rangle^2)$, where $\hat{H}$ is the generator of the unitary transformation and the average is taken with respect to the initial quantum state \cite{BRAUNSTEIN1996135}. Therefore, the quantum Fisher information for a coherent state with a mean photon number $N$ in the spatial mode of Eq.\eqref{Hermite-mode} is given by
\begin{equation}\label{quantum-Fisher-lateral}
    \mathcal{F}=4N\langle HG_{m,0}|\hat{p}_x^2|HG_{m,0}\rangle=\frac{4N(2m+1)}{w^2}\,.
\end{equation}
from which we can evidence the metrological enhancement provided by the higher order modes once the uncertainty $\delta d$ is reduced by a $\sqrt{2m+1}$ factor.

By measuring the probabilities of detecting the photon state in the original mode $HG_{m,0}$ ($P$) and in the orthogonal modes $HG_{m+1,0}$ and $HG_{m-1,0}$ ($P'$), we get the following probabilities from Eq.\eqref{mode-expansion-lateral}: $P=1-(2m+1)d^2/w^2$ and $P'=(2m+1)d^2/w^2$. The Fisher information arising from this (binary) measurement with $N$ uncorrelated photons is then
\begin{equation}
    F(d)=\frac{4N(2m+1)}{w^2}+\mathcal{O}(d^2/w^2)\,,
\end{equation}
showing that measuring the crosstalk of an HG modes with its neighbouring modes is a quantum-optimal way of sensing small displacements. 

Similarly for the Laguerre-Gauss modes, the quantum Fisher information is given by \cite{supplemental-material}
\begin{equation}\label{quantum-Fisher-lateral-LG}
    \mathcal{F}=4N\langle LG_{m,n}|\hat{p}_x^2|LG_{m,n}\rangle=\frac{4N(m+n+1)}{w^2}\,.
\end{equation}

Analogously to the previous case, we can expand the displaced mode in the original mode basis \cite{supplemental-material}. The probabilities associated with measuring the original state ($LG_{m,n}$) and the (orthogonally) populated ones ($LG_{m+1,n}$, $LG_{m,n+1}$, $LG_{m-1,n}$ and $LG_{m,n-1}$) are given by $P=1-(m+n+1)d^2/w^2$ and $P'=(m+n+1)d^2/w^2$ \cite{supplemental-material}. The corresponding Fisher information with $N$ uncorrelated photons is then
\begin{equation}
    F(d)=\frac{4N(m+n+1)}{w^2}+\mathcal{O}(d^2/w^2)\,,
\end{equation}
showing once again that the proposed measurement protocol is optimal.

Similar calculations for the case of angular displacements are shown in the supplementary material \cite{supplemental-material} for both HG and LG beams, leading to the quantum-optimal Fisher information in both cases. This is not surprising since the lateral displacement can be switched to a tilt by a Fourier transformation implemented on a paraxial beam.

\begin{figure}
    \centering
    \includegraphics[scale=1.0]{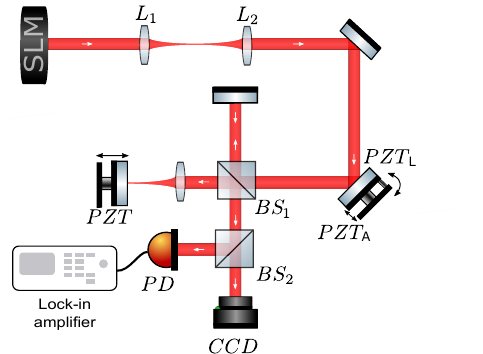}
    \caption{Experimental setup. A spatial light modulator (SLM) prepares the desired spatial mode. The signal from the photo-detector (PD) is demodulated in a lock-in amplifier (Liquid Instruments, Moku:Pro).}
    \label{fig:exp-setup}
\end{figure}

In order to realize such measurements, we developed an experimental setup based on a Michelson interferometer, in which one of the arms was modified with a lens, as shown in Figure \ref{fig:exp-setup}. The collimated beam is split by a beam splitter (BS) into two distinct paths. One of the paths consists of a 4f system (2f on the way forward and 2f on the way back) and a mirror positioned at the beam's focus, so that the beam returns exactly along the same incoming path. This 4f system implements the transformation \(\hat{r} \rightarrow -\hat{r}\) on the input field. This image inversion is equivalent to adding an extra $\pi$ phase to all odd modes in the input field while maintaining the even modes intact, as a consequence of the imaging system. Hence, when the longitudinal phase difference in the two arms are equivalent to zero ($\varphi=0$), the bright port for the odd modes corresponds to the dark port for the even modes, i.e., the modes are separated by their parity. The probabilities $P$ and $P'$ are obtained by monitoring the output ports of the interferometer in the situation where $\varphi=0$. 

It is straightforward to show that for a horizontally polarized beam, the field in one output port on the interferometer is given by $E=E_e\sin\varphi/2+E_o\cos\varphi/2$, where $E_{e(o)}$ is the even (odd) spatial mode of the input beam, and $\varphi$ is the longitudinal phase difference in the interferometer. Since the optical power $I$ is proportional $|E|^2$, we have for the output power
\begin{equation}\label{interferometer-output}
    I=\frac{I_{tot}}{2}\left(1+\nu\cos\varphi\right)\,,
\end{equation}
where $I_{tot}$ is the total input optical power, and $\nu=|(|E_o|^2-|E_e|^2)|/(|E_o|^2+|E_e|^2)=(|I_o-I_e|)/(I_o+I_e)$ is the visibility of the interferometer, which reveals the contrast between the different parity modes. For an input mode with well-defined parity ($HG_{m,0}$ or $LG_{m,n}$) the visibility is equal to $1$. However, for a displaced beam with respect to the optical axis of the interferometer, the visibility will be less than 1, since it presents a small component in the opposite parity. Since the optical power in a given mode is proportional to the probabilities of individual photons, we can write that $\nu=1-2P'$, such that
\begin{eqnarray}
    \nu^{(HG)}&=&1-2(2m+1)d^2/w^2 \,,\label{visibility-lateral-HG}\\ 
    \nu^{(LG)}&=&1-2(m+n+1)d^2/w^2\,, \label{visibility-lateral-LG}
\end{eqnarray}
for the angular displacement for HG and LG modes, respectively.

\textit{Experiment.}---The experimental setup is depicted in Fig.\eqref{fig:exp-setup}. A spatial light modulator (Hamamatsu LCOS-SLM X10468) is used to generate the spatial modes from a $633$ nm He-Ne laser beam. We utilized the repository \cite{Gil_de_Oliveira_slmcontrol_2024-1} to generate the necessary holograms for producing the transverse modes. The generated beam is collimated by the lenses $L_1$ and $L_2$ to a width of $w=590\,\mu m$. Two piezoelectric actuators are attached to the mirror mount just before the interferometer: the $PZT_L$ is attached to the back of the mirror to ensure a pure lateral displacement of the beam; the $PZT_A$ a screw actuator (Thorlabs Polaris-P20A) installed on the mirror mount barrel to ensure angular displacement.  On one of the interferometer mirrors, we coupled a third piezoelectric actuator ($PZT$) which allows us to fine-tune the relative phase $\varphi$ of the interferometer. The output ports of the interferometer are monitored by a CCD camera, to visually check the interference quality, and a photo-detector (Thorlabs PDA100A2).

We choose a standard and convenient way to attest to the metrological gain of our method, applying an external modulation $\Omega$ to the displacement parameter such that $d\to d_0+d\sin(\Omega t)$, where $d_0$ account for an unavoidable experimental offset, and substituting Eqs.\eqref{visibility-lateral-HG} and \eqref{visibility-lateral-LG} into \eqref{interferometer-output} with their respective modulations, we end up with 
\begin{align}
 I^{(HG)}=\frac{I_{tot}}{2}\Big[\Gamma_0-4(2m+1)\frac{d_0d}{w^2}\cos\varphi\sin(\Omega t) \nonumber\\ +(2m+1)\frac{d^2}{w^2}\cos\varphi\cos(2\Omega t)\Big]\,, \label{output-modulation-HG} \\
 I^{(LG)}=\frac{I_{tot}}{2}\Big[\Gamma'_0-4(m+n+1)\frac{d_0d}{w^2}\cos\varphi\sin(\Omega t) \nonumber\\ +(m+n+1)\frac{d^2}{w^2}\cos\varphi\cos(2\Omega t)\Big]\,. \label{output-modulation-LG}
\end{align}

Therefore, we see that there are two spectral components ($\Omega$ and $2\Omega$) that carry information about the displacement, which will be maximized at $\varphi=\{0,\pi\}$, precisely when the interferometer acts as a parity selector. The frequency $2\Omega$ will be most relevant, since its magnitude depends only on the amplitude of the modulation, unlike the frequency component $\Omega$. Moreover, they exhibit the magnification factor of \{$(2m+1)$,$(m+n+1)$\} that characterizes the metrological enhancement present in Eqs.\eqref{quantum-Fisher-lateral} and \eqref{quantum-Fisher-lateral-LG}.

Initially, we align the interferometer with the incoming laser beam in order to ensure the highest possible visibility, and we consider this as our starting point. The visibility we obtained for the Gaussian ($HG_{0,0}$) mode is $97\%$. After the initial alignment procedures, we apply the displacement (lateral or angular) modulation with peak-to-peak drive voltage of $0.8$ V at a frequency $\Omega=650$ Hz, well bellow the mechanical resonance of the piezos. The measured signal is demodulated in a lock-in amplifier at frequency $2\Omega$ to measure the amplitude of the applied modulation. As seen from Eqs.\eqref{output-modulation-HG} and \eqref{output-modulation-LG} the demodulated signal is proportional to $\cos\varphi$, which implies that it is maximized at the constructive (destructive) output port, when $\varphi=0$ ($\varphi=\pi$). Therefore, we experimentally drive the interferometer piezoactuator (PZT) in order to maximize the demodulated signal for each incoming transverse mode. We verified that the phase remains stable during the measurements for each transverse mode.

\begin{figure}
    \includegraphics[scale=0.42]{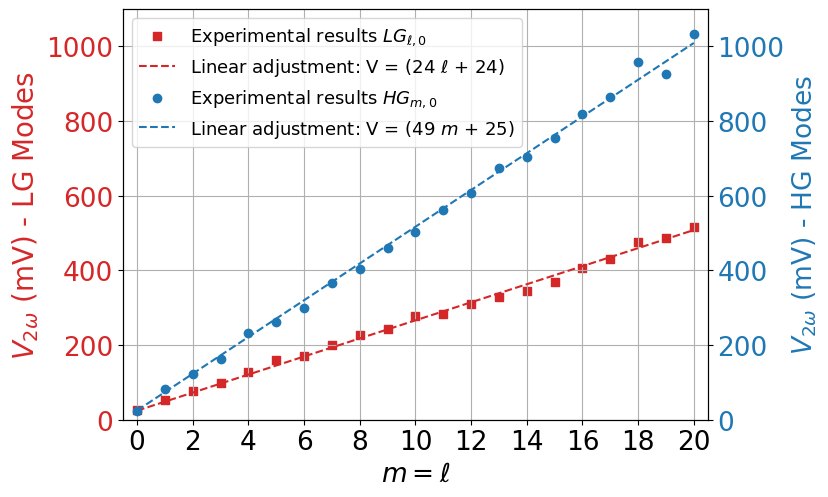}
    \caption{Results for lateral displacements. Blue dots represent measurements using $HG_{m,0}$ modes, while red dots represent results using $LG_{\ell,0}$ modes. The red and blue lines indicate linear fits to the respective data. Error bars are small and therefore not shown in the figure.
} 
    \label{fig:experimental-result}
\end{figure}

The experimental results for the lateral displacement for the Hermite-Gauss modes ($HG_{m,0}$) are depicted in Fig.\eqref{fig:experimental-result}. It shows the amplitude of the demodulated signal at frequency $2\Omega$ as a function of the input spatial mode for a given modulation amplitude. The gain in the lock-in amplifier was set to $80$ dB. In order to capture the dependence on the mode order \(m\), we fixed the optical power at \(P = 1.0\,\mu W\). For greater clarity, detailed tables with specific values for each set of results are presented in the supplementary material \cite{supplemental-material}.

We note from Eq.\eqref{output-modulation-HG}, that the amplitude at frequency $2\Omega$ is linearly proportional to the mode order $m$, as $V_{2\Omega}=V_0(2m+1)$, where $V_0$ is the signal value for the Gaussian mode. Thus, to check the dataset trend, we performed a linear fit with $V_0$ as the free parameter. The obtained value $25$ mV agrees with the experimental value $(25\pm 2)$ mV, where the error was taken from the random fluctuations over repeated measurements. This result shows the magnification coming from the Hermite-Gauss mode order $m$, which amplifies the signal but not the noise. The overall effect is the increase in the SNR, which allows for detection of smaller tilt values. The results concerning the angular displacement for the Hermite-Gauss modes are shown in the supplementary material \cite{supplemental-material}.

In the same Figure \ref{fig:experimental-result}, we show the results Laguerre-Gaussian modes with $p = 0$, which  corresponds $m = l$ and $n = 0$, where the displacement is applied in the same $\hat{x}$ direction as before. As shown in Eq. \eqref{output-modulation-LG}, the response at frequency $2\Omega$ is linearly proportional to $l + 1$, yielding $V_{2\Omega} = V_0(l + 1)$. A linear fit with $V_0$ as a free parameter gives a value of $24$ mV, in agreement with the experimental measurement of ($26 \pm 2$) mV. In the supplementary material \cite{supplemental-material}, we illustrate the case where $l = 0$, leading to $m = n = p$. 

\begin{figure}
    \includegraphics[scale=.29]{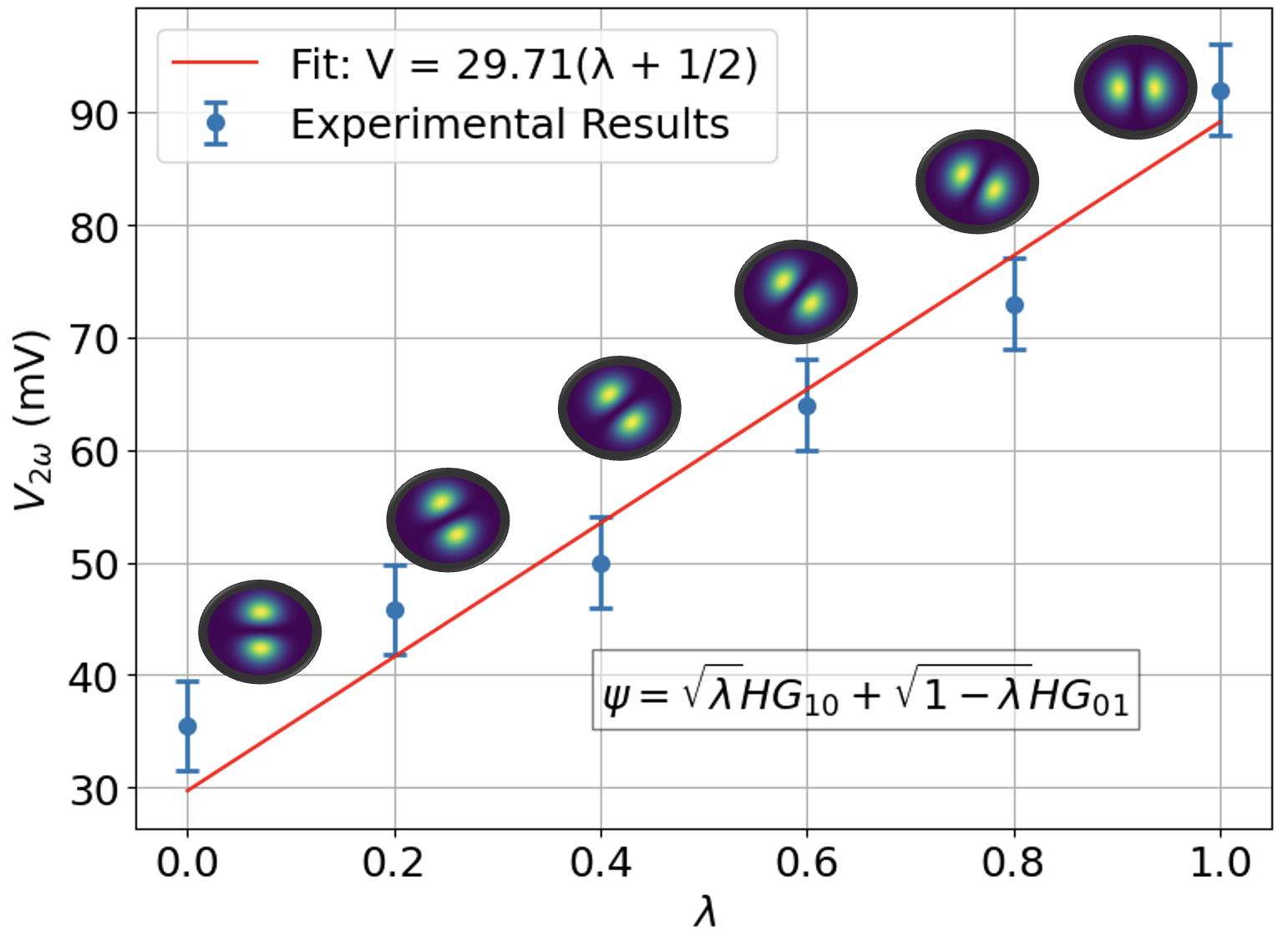}
    \caption{A lateral displacement in the $\hat{x}$ direction is applied to the superposition $\psi$. The figure presents experimental results showing how the signal amplitude varies with $\lambda$, clearly highlighting an advantage when it takes the $HG_{10}$ form compared to $HG_{01}$.}
    \label{fig:experimental-result-sup}
\end{figure}

Due to the cylindrical symmetry of the LG modes, the displacement can take any direction in the transverse plane without affecting the efficiency of the protocol. In contrast, HG modes are more efficient only when the spatial modulation coincides with the displacement direction. Figure \eqref{fig:experimental-result-sup} illustrates this effect through a superposition of HG modes. By analyzing the resulting signal-to-noise ratio (SNR), it is observed that the displacement information increases progressively as the nodal line of the beam becomes more perpendicular to the direction of the displacement. The point at $\lambda = 1$ corresponds to the Hermite-Gaussian mode $HG_{10}$, where the highest metrological gain is observed, as expected.

Thus, the LG modes allow for the calculation of the magnitude of the displacement in any direction but do not provide information about its direction. The demonstrated effect is that the HG modes can be used to determine not only the magnitude but also the direction in which the displacement is occurring.

\textit{Conclusion.}---We have developed an interferometric technique to measure lateral and angular displacements of an optical beam with optimal sensitivity. The technique relies on the enhanced metrological performance provided by the high-order spatial modes, coming from their well-defined inversion symmetry. Notably, for both types of displacements, we achieved a 41-fold improvement in the signal-to-noise ratio compared to the fundamental Gaussian mode when using high-order Hermite-Gaussian modes ($m=20$). Essentially, the method enables the use of high-order structured modes without requiring high-order local oscillators in superposition and homodyne detection. Comparatively, achieving the metrological gain of the $HG_{20,0}$ mode with squeezed light in lower-order modes, such as $HG_{1,0}$, would demand $16$ dB of compression, a value not reached so far, to the best of our knowledge. A 21-fold improvement factor was obtained for the case of Laguerre-Gaussian modes.

Finally, our technique offers a practical solution for real-world scenarios: in a non-modulated regime, one can track the interferometer’s visibility (Eqs.\eqref{visibility-lateral-HG} and \eqref{visibility-lateral-LG}) to estimate the displacement parameter in real time. We confirm, for the first time, that Laguerre-Gaussian modes enable optimal small-displacement detection, in line with the quantum Fisher information. Moreover, by exploiting the asymmetry of Hermite-Gaussian modes, we can discern the direction of unknown displacements, thus extending this method to any system featuring modes with well-defined parity.

\textit{Acknowledgements.}---The authors would like to acknowledge the financial support from the Brazilian agencies: Conselho Nacional de Desenvolvimento Científico e Tecnológico (CNPq), Coordenação de Aperfeiçoamento de Pessoal de Nível Superior (CAPES), Fundação Carlos Chagas Filho de Amparo à Pesquisa do Estado do Rio de Janeiro (FAPERJ) and Instituto Nacional de Ciência e Tecnologia de Informação Quântica (INCT-IQ 465469/2014-0) and Fundação de Amparo à Pesquisa do Estado de São Paulo (FAPESP, grants 2021/06823-5 and 2022/15036-0).

\bibliography{references}

\clearpage
\onecolumngrid                    

\renewcommand{\thefigure}{S\arabic{figure}}
\renewcommand{\thetable}{S\arabic{table}}
\setcounter{figure}{0}
\setcounter{table}{0}

\begin{center}
  \textbf{\Large Supplemental Material}\\[4pt]
\end{center}
\vspace{12pt}

Here we outline the calculations needed to support the results in the Theory section of the manuscript.

The Hermite and Laguerre-Gaussian modes can be described by an operator formalism, borrowed from quantum mechanics. Let $\hat{x},\hat{y}$ be operators acting on the Hilbert space formed by the set of complex valued square integrable functions in $\mathbb{R}^2$, which are defined by the expressions $\hat{x} f(x,y) = x f(x,y)$ and $\hat{y} f(x,y) = y f(x,y)$. We also introduce the momentum operators $\hat{p}_x = - i \partial_x$ and $\hat{p}_{y} = -i \partial_y$. These operators follow the standard commutation relations $\left[ \hat{x}, \hat{p}_x \right] = \left[ \hat{y}, \hat{p}_y \right] = i$. Now, we define the lowering operator
\begin{equation}
    \hat{a}_x = \frac{\hat{x}}{w} + \frac{i w \hat{p}_x}{2},
\end{equation}
where $w$ is a parameter with dimensions of length that corresponds to the waist of our modes. 

A Hermite-Gaussian mode can then be written as
\begin{equation}
    \ket{HG_{mn}} = \left( a^\dagger_x \right)^m \left( a^\dagger_y \right)^n \ket{0}
\end{equation}
where $\ket{0}$ is the fundamental Gaussian mode that satisfies $\hat{a}_x \ket{0} = \hat{a}_y \ket{0} = 0$. Using the the standard properties of the raising and lowering operators, we trivially get that
\begin{equation}
    \begin{aligned}
        \hat{x}\ket{HG_{mn}} &= \frac{w}{2} \left( \hat{a}^\dagger_x + \hat{a}_x \right) \ket{HG_{mn}} \\ &= \frac{w}{2} \left( \sqrt{m+1} \ket{HG_{m+1,n}} + \sqrt{m} \ket{HG_{m-1,n}}  \right)
    \end{aligned}
\end{equation}
and, consequently,
\begin{equation}
    \expval{\hat{x}^2}{HG_{mn}} = \frac{w^2}{4} (2m+1)
\end{equation}

For the momentum operator, we get
\begin{equation}
    \begin{aligned}
        \hat{p}_x\ket{HG_{mn}} &= \frac{i}{w} \left( \hat{a}^\dagger_x - \hat{a}_x \right) \ket{HG_{mn}} \\ &= \frac{i}{w} \left( \sqrt{m+1} \ket{HG_{m+1,n}} - \sqrt{m} \ket{HG_{m-1,n}}  \right)
    \end{aligned}
\end{equation}
and
\begin{equation}
    \expval{\hat{p}^2_x}{HG_{mn}} = \frac{1}{w^2} (2m+1)
\end{equation}

The Laguerre-Gaussian modes follow a similar construction based on the new ladder operators
\begin{equation}
    \hat{a}^\dagger_{\pm} = \frac{\hat{a}^\dagger_x \pm i\hat{a}^\dagger_y}{\sqrt{2}}.
\end{equation}
The modes are then defines by
\begin{equation}
    \ket{LG_{mn}} = \left( a^\dagger_+ \right)^m \left( a^\dagger_- \right)^n \ket{0}
\end{equation}
where $m,n$ is an alternative pair of labels for the modes, which are related to the radial index $p = \min(m,n)$ and the topological charge $l = m-n$. These relations may be inverted, from where we obtain $m = p + (\abs{l}+l)/2; \ n = p + (\abs{l}-l)/2$.

We may write
\begin{equation}
    \hat{a}_x = \frac{\hat{a}_+ + \hat{a}_-}{\sqrt{2}};  \ \ \ \hat{a}_y = \frac{i\left( \hat{a}_+ - \hat{a}_- \right)}{\sqrt{2}}  
\end{equation}
so that
\begin{equation}
    \begin{aligned}
        \hat{x} \ket{LG_{mn}} = \frac{w}{2\sqrt{2}} &\left(  \sqrt{m+1} \ket{LG_{m+1,n}} + \sqrt{m} \ket{LG_{m-1,n}} \right.  \\ & \left. + \sqrt{n+1} \ket{LG_{m,n+1}} + \sqrt{n}\ket{LG_{m,n-1}} \right)
    \end{aligned}
\end{equation}

\begin{equation}
    \expval{\hat{x}^2}{LG_{mn}} = \frac{w^2}{4}(m+n+1)
\end{equation}

\begin{equation}
    \begin{aligned}
        \hat{p}_x \ket{LG_{mn}} = \frac{i}{\sqrt{2}w} &\left(  \sqrt{m+1} \ket{LG_{m+1,n}} - \sqrt{m} \ket{LG_{m-1,n}} \right.  \\ & \left. + \sqrt{n+1} \ket{LG_{m,n+1}} - \sqrt{n}\ket{LG_{m,n-1}} \right)
    \end{aligned}
\end{equation}

\begin{equation}
    \expval{\hat{p}_x^2}{LG_{mn}} = \frac{1}{w^2}(m+n+1)
\end{equation}

With these results at hand, we can now describe the case of angular displacement for a structured light mode. An angular displacement $g$ in the laser beam can be seen as an unitary operator $\exp(-ig\hat{x})$, where $\hat{x}$ is the transverse position operator and $g$ is the transverse momentum acquired by the beam. Let us first consider the case of a displaced Hermite-Gauss beam.

The quantum Fisher information for a coherent state with a mean photon number $N$ is given by
\begin{equation}\label{quantum-Fisher-tilt}
    \mathcal{F}=4N\langle HG_{m,0}|\hat{x}^2|HG_{m,0}\rangle=N(2m+1)w^2\,.
\end{equation}

Under a small tilt ($gw\ll 1$), the beam can be expanded in the original $HG$ basis as
\begin{align}\label{mode-expansion-tilt}
    &e^{-ig\hat{x}}\ket{HG_{m,0}}\approx (1-ig\hat{x})\ket{HG_{m,0}} = \ket{HG_{m,0}}\nonumber\\
    &+\frac{igw}{2}\left(\sqrt{m+1}\,\ket{HG_{m+1,0}}+\sqrt{m}\,\ket{HG_{m-1,0}}\right)\,
\end{align}
for $m\geq 1$ and neglecting second-order terms in the series expansion.

By measuring the probabilities of detecting the photon state in the original mode $HG_{m,0}$ ($P$) or in the populated modes $HG_{m+1,0}$ and $HG_{m-1,0}$ ($P'$), we get the following probabilities from Eq.\eqref{mode-expansion-tilt}: $P=1-(2m+1)g^2w^2/4$ and $P'=(2m+1)g^2w^2/4$. The Fisher information arising from this (binary) measurement with $N$ uncorrelated photons from a coherent state is given by
\begin{equation}
    F(g)=N(2m+1)w^2+\mathcal{O}(g^2w^2)\,.
\end{equation}
Hence, the proposed measurement is optimal under the small tilt approximation.




Consider now the case of a small tilt in a Laguerre-Gauss beam. The quantum Fisher information for this process is given by
\begin{equation}\label{quantum-Fisher-LG}
     \mathcal{F}=N\langle LG_{m,n}|\hat{x}^2|LG_{m,n}\rangle=N(m+n+1)w^2\,.
\end{equation}

Analogously, expanding the displaced mode in the original mode basis, we have that
\begin{align}\label{mode-expansion-LG}
     &e^{-ig\hat{x}}\ket{LG_{m,n}}\approx (1-ig\hat{x})\ket{LG_{m,n}}= \ket{LG_{m,n}} \nonumber\\
     &-\frac{igw}{2\sqrt{2}}\Big(\sqrt{m+1}\,\ket{LG_{m+1,n}}+\sqrt{m}\,\ket{LG_{m-1,n}} \nonumber\\
     &+\sqrt{n+1}\,\ket{LG_{m,n+1}}+\sqrt{n}\,\ket{LG_{m,n-1}}\Big)\,,
\end{align}
and the probabilities associated with measuring the photon in the original state ($LG_{m,n}$) and the populated ones ($LG_{m+1,n}$, $LG_{m-1,n}$, $LG_{m,n+1}$ and $LG_{m,n-1}$) are given by $P=1-(m+n+1)g^2w^2/4$ and $P'=(m+n+1)g^2w^2/4$. The corresponding Fisher information with $N$ uncorrelated photons is then
\begin{equation}\label{eq-fisher-interferometer-LG}
     F(g)=N(m+n+1)w^2+\mathcal{O}(g^2w^2)\,,
\end{equation}
showing once again that the proposed measurement protocol is optimal.

Finally, the case of a lateral displacement of a Laguerre-Gauss beam is accounted by the quantum Fisher information:
\begin{equation}\label{quantum-Fisher-lateral-LG}
    \mathcal{F}=4N\langle LG_{m,n}|\hat{p}_x^2|LG_{m,n}\rangle=\frac{4N(m+n+1)}{w^2}\,.
\end{equation}

The first-order expansion of the displaced Laguerre-Gauss mode is given by:
\begin{align}\label{mode-expansion-d-LG}
     &e^{-id\hat{p}_x}\ket{LG_{m,n}}\approx (1-id\hat{p}_x)\ket{LG_{m,n}}= \ket{LG_{m,n}} \nonumber\\
     &+\frac{d}{\sqrt{2}w} \left(  \sqrt{m+1} \ket{LG_{m+1,n}} - \sqrt{m} \ket{LG_{m-1,n}} \right.  \nonumber\\
     & \left. + \sqrt{n+1} \ket{LG_{m,n+1}} - \sqrt{n}\ket{LG_{m,n-1}} \right)\,,
\end{align}
such that the probabilities associated with measuring the original state ($LG_{m,n}$) and the populated ones ($LG_{m+1,n}$, $LG_{m,n+1}$, $LG_{m-1,n}$ and $LG_{m,n-1}$) are given by $P=1-(m+n+1)d^2/w^2$ and $P'=(m+n+1)d^2/w^2$. The corresponding Fisher information with $N$ uncorrelated photons is then
\begin{equation}
    F(d)=\frac{4N(m+n+1)}{w^2}+\mathcal{O}(d^2/w^2)\,,
\end{equation}
showing once again that the proposed measurement protocol is optimal.

\begin{figure}[h]
    \centering
    \includegraphics[width=1\linewidth]{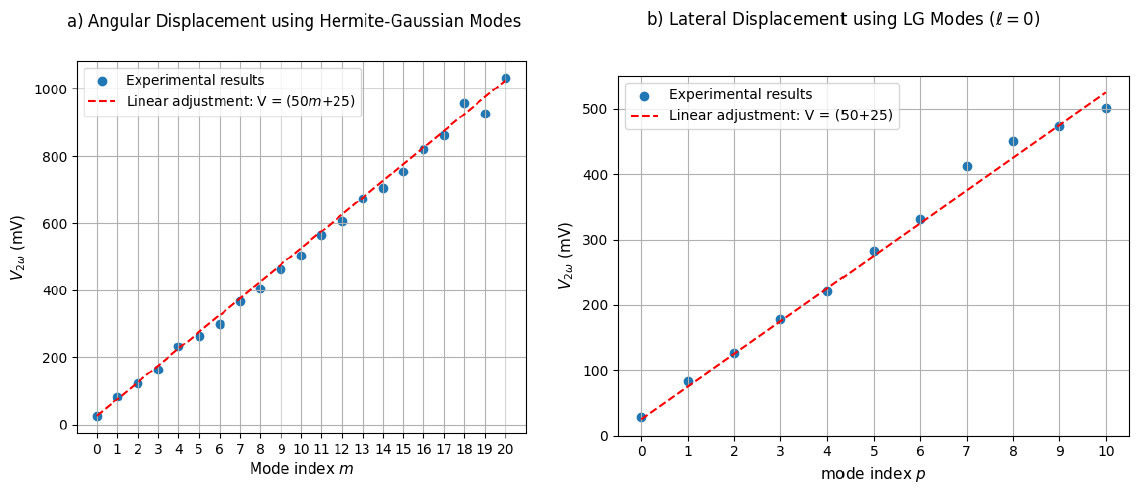}
    \caption{Figure (a) presents the results for angular displacement using Hermite-Gaussian modes $HG_{m,0}$, while figure (b) shows the results for Laguerre-Gaussian modes with $l = 0$ and $m = n = p$, for lateral displacement. In both figures, the blue dots represent the experimental measurements, whereas the red lines correspond to linear fits to the obtained data. The error bars are small and, therefore, are not displayed in the figures.}
    \label{fig:resulta_sm}
\end{figure}

 The results the angular displacement for the Hermite-Gauss modes are shown in the \eqref{fig:resulta_sm} (a). In order to capture the dependence on the mode order \(m\), we fixed the optical power at \(P = 2.0\,\mu W\), except for modes with order higher than \(m = 10\), which were measured with \(P = 1.1\,\mu W\) to avoid the \(1\,V\) saturation in the lock-in amplifier. We carefully rescaled these values owing to its proportionality with the input optical power. This procedure was applied to all results, and we have provided detailed tables below with specific values for each set of results. The fitting result for the $V_0$ parameter was $50$ mV, which matches with the experimental value $(48 \pm 2$) mV, showing good
agreement with the theory.

In Figure \ref{fig:resulta_sm} (b), we illustrate the case where $l = 0$, leading to $m = n = p$ for lateral displacement. For this configuration, the linear fit is performed with $V_{2\Omega} = V_0(2p + 1)$, resulting in a fitted value of $25$ mV, compared to the experimental result of ($28 \pm 2$) mV.

In the tables below, we present the experimental results, with the error bars, obtained directly from the Lock-in amplifier. All modes were carefully calibrated to have the same power $P$ ($\mu$W), as indicated in the tables, ensuring data consistency. Due to the saturation of the Lock-in amplifier at 1.1 V, it was necessary to recalibrate to a lower power for the HG$_{m,0}$ modes in angular displacement with $m$ greater than 10. Therefore, we rescaled the obtained values, considering they are proportional to the input power.

\onecolumngrid

\begin{table}[h!]
\centering
\captionsetup{labelformat=empty} 
\caption{\large Using Hermite-Gaussian modes}
\vspace{0.3cm} 

\captionsetup{labelformat=default}
\setcounter{table}{0}

\begin{minipage}{0.45\linewidth}
    \centering
    \label{table:hermite_gaussian_lateral}
    \large 
    \begin{tabular}{|c|c|c|}
    \hline
    \text{$m$} & $V_{2\Omega}$ (mW) & P ($\mu$W) \\
    \hline
    0  & $25 \pm 2$   & \\
    1  & $83 \pm 2$   & \\
    2  & $124 \pm 2$  & \\
    3  & $165 \pm 2$  & \\
    4  & $234 \pm 2$  & \\
    5  & $265 \pm 2$  & \\
    6  & $299 \pm 2$  & \\
    7  & $367 \pm 2$  & \\
    8  & $407 \pm 4$  & \\
    9  & $464 \pm 4$  & \\
    10 & $507 \pm 4$  & $2.0\pm 0.1$ \\
    11 & $566 \pm 4$  & \\
    12 & $609 \pm 4$  & \\
    13 & $676 \pm 4$  & \\
    14 & $706 \pm 4$  & \\
    15 & $758 \pm 4$  & \\
    16 & $824 \pm 7$  & \\
    17 & $869 \pm 8$  & \\
    18 & $965 \pm 8$  & \\
    19 & $932 \pm 8$  & \\
    20 & $1038 \pm 8$ & \\
    \hline
    \end{tabular}
    \caption{Lateral displacement} 
\end{minipage}%
\hspace{1cm} 
\begin{minipage}{0.45\linewidth}
    \centering
    \large 
    \label{table:hermite_gaussian_angular}
    \begin{tabular}{|c|c|c|}
    \hline
    $m$ & $V_{2\Omega}$ (mW) & P ($\mu$W) \\
    \hline
    0  & $48 \pm 4$    & \\
    1  & $136 \pm 4$   & \\
    2  & $249 \pm 4$   & \\
    3  & $353 \pm 4$   & \\
    4  & $410 \pm 4$   & \\
    5  & $525 \pm 4$   & $2.0\pm0.1$ \\
    6  & $647 \pm 6$   & \\
    7  & $799 \pm 8$   & \\
    8  & $860 \pm 8$  & \\
    9  & $960 \pm 10$   & \\
    10 & $1062 \pm 10$  &  \\
    \cline{1-3} 
    11 & $593 \pm 4$  & \\
    12 & $619 \pm 4$   & \\
    13 & $688 \pm 4$   & \\
    14 & $772 \pm 6$   & \\
    15 & $796 \pm 6$   & $1.0\pm 0.1$ \\
    16 & $815 \pm 8$   & \\
    17 & $861 \pm 8$   & \\
    18 & $918 \pm 8$   & \\
    19 & $946 \pm 10$   & \\
    20 & $1066 \pm 10$  & \\
    \hline
    \end{tabular}
    \caption{Angular displacement} 
\end{minipage}
\end{table}

\begin{table}[h!]
\centering
\captionsetup{labelformat=empty} 
\caption{\large Using Laguerre-Gaussian modes}
\vspace{0.3cm} 

\captionsetup{labelformat=default}
\setcounter{table}{2}

\begin{minipage}{0.45\linewidth}
    \centering
    \large 
    \label{table:Laguerre_gaussian_l=0}
    \begin{tabular}{|c|c|c|}
    \hline
    $p$ & $V_{2\Omega}$ (mW) & P ($\mu$W) \\
    \hline
    0  & $29 \pm 4$   & \\
    1  & $83 \pm 4$   & \\
    2  & $126 \pm 4$  & \\
    3  & $179 \pm 4$  & \\
    4  & $221 \pm 6$  & \\
    5  & $283 \pm 6$  & $1.0\pm0.1$ \\
    6  & $331 \pm 6$  & \\
    7  & $412 \pm 6$  & \\
    8  & $449 \pm 8$  & \\
    9  & $473 \pm 8$  & \\
    10 & $501 \pm 8$  &  \\
    \hline
    \end{tabular}
    \caption{With $\ell = 0$} 
\end{minipage}%
\hspace{1cm} 
\begin{minipage}{0.45\linewidth}
    \centering
    \large 
    \label{table:Laguerre_gaussian_p=0}
    \begin{tabular}{|c|c|c|}
    \hline
    $\ell$ & $V_{2\Omega}$ (mW) & P ($\mu$W) \\
    \hline
    0  & $26 \pm 4$    & \\
    1  & $54 \pm 4$   & \\
    2  & $77 \pm 4$   & \\
    3  & $99 \pm 4$   & \\
    4  & $127 \pm 4$   & \\
    5  & $159 \pm 4$   &  \\
    6  & $170 \pm 6$   & \\
    7  & $201 \pm 6$   & \\
    8  & $228 \pm 6$  & \\
    9  & $244 \pm 6$   & \\
    10 & $278 \pm 6$  & $1.0\pm0.1$ \\
    11 & $284 \pm 8$  & \\
    12 & $309 \pm 8$   & \\
    13 & $328 \pm 8$   & \\
    14 & $345 \pm 8$   & \\
    15 & $370 \pm 8$   &  \\
    16 & $406 \pm 10$   & \\
    17 & $430 \pm 10$   & \\
    18 & $477 \pm 10$   & \\
    19 & $487 \pm 10$   & \\
    20 & $516 \pm 10$  & \\
    \hline
    \end{tabular}
    \caption{With $p = 0$} 
\end{minipage}
\end{table}

\end{document}